\documentclass[useAMS,usenatbib]{mnras}
\usepackage{graphics,epsfig,psfig}
\usepackage[normalem]{ulem}
\usepackage{float}
\usepackage{subfig}
\usepackage{graphicx}
\usepackage[]{inputenc,amssymb}
\usepackage{gensymb}
\usepackage{soul}
\usepackage{todonotes}

\def \be{\begin{equation}}
\def \ee{\end{equation}}
\def \bea{\begin{eqnarray}}
\def \eea{\end{eqnarray}}

\hypersetup{colorlinks=true,allcolors=magenta}

\definecolor{webgreen}{rgb}{0,.5,0}
\definecolor{webbrown}{rgb}{.6,0,0}
\definecolor{cadmiumgreen}{rgb}{0.0, 0.42, 0.24}

\title[Testing GR from dark standard sirens]{Testing the general theory of relativity using gravitational wave propagation from dark standard sirens}
\author[Mukherjee, Wandelt, \& Silk]{Suvodip Mukherjee$^{1,2,3}$\thanks{s.mukherjee@uva.nl, mukherje@iap.fr}, Benjamin D. Wandelt$^{4, 5, 6}$ \thanks{wandelt@iap.fr} \& Joseph Silk$^{4, 5, 7, 8}$ \thanks{j.silk1@physics.ox.ac.uk}\\
$^{1}$ Gravitation Astroparticle Physics Amsterdam (GRAPPA),
Anton Pannekoek Institute for Astronomy and Institute for Physics,\\
University of Amsterdam, Science Park 904, 1090 GL Amsterdam, The Netherlands\\
$^2$ Institute Lorentz, Leiden University, PO Box 9506, Leiden 2300 RA, The Netherlands\\
$^3$Delta Institute for Theoretical Physics, Science Park 904, 1090 GL Amsterdam, The Netherlands\\
$^{4}$ Institut d'Astrophysique de Paris, UMR 7095, CNRS, Sorbonne Universit\'e, 98bis Boulevard Arago, 75014 Paris, France\\
$^{5}$ Sorbonne Universit\'e, Institut Lagrange de Paris,  98 bis Boulevard Arago, 75014 Paris, France\\
$^{6}$ Center for Computational Astrophysics, Flatiron Institute, 162 5th Avenue, 10010, New York, NY, USA\\
$^{7}$ The Johns Hopkins University, Department of Physics \& Astronomy, 3400 N. Charles Street, Baltimore, MD 21218, USA\\
$^{8}$ Beecroft Institute for Cosmology and Particle Astrophysics, University of Oxford, Keble Road, Oxford OX1 3RH, UK\\
}
\pubyear{2020}
\begin{document}
\label{firstpage}
\pagerange{\pageref{firstpage}--\pageref{lastpage}}
\maketitle

\label{firstpage}

\begin{abstract}
 Alternative theories of gravity predict modifications in the propagation of gravitational waves (GW) through space-time. One of the smoking-gun predictions of such theories is the change in the GW luminosity distance to GW sources as a function of redshift relative to the electromagnetic (EM) luminosity distance expected from  EM probes. 
We propose a multi-messenger test of the theory of general relativity from the propagation of gravitational waves by combining EM and GW  observations to resolve these issues from GW sources without EM counterparts (which are also referred to as  dark standard sirens). By using the relation between the geometric distances accessible from  baryon acoustic oscillation measurements, and luminosity distance measurements from the GW sources, we can measure any deviation from the general theory of relativity via the  GW sources of unknown redshift that will be detectable by networks of GW detectors such as LIGO, Virgo, and KAGRA. Using this technique, the fiducial value of the frictional term can be measured to a precision $\Xi_0=0.98^{+0.04}_{-0.23}$ after marginalizing over  redshift dependence, cosmological parameters, and  GW bias parameters with $\sim 3500$ dark standard sirens of masses $30\,\rm M_\odot$ each distributed up to redshift $z=0.5$. For a fixed redshift dependence, a value of $\Xi_0=0.99^{+0.02}_{-0.02}$ can be measured with a similar  number of dark sirens. 
Application of our methodology to the far more numerous dark standard sirens detectable with  next generation GW detectors, such as LISA, Einstein Telescope and Cosmic Explorer, will allow  achievement of  higher accuracy than possible from  use of  bright standard sirens.
\end{abstract}

\begin{keywords} 
gravitational waves, large-scale structure of Universe
\end{keywords}

\section{Introduction}
 The general theory of relativity predicts a unique geodesic for both electromagnetic wave  {(EM)} and gravitational wave (GW) signals. However several alternative theories of gravity predict deviations from the general theory of relativity by having a difference in the speed of propagation between GW and EM signals, due to the non-zero mass of the graviton, the running of the effective Planck mass (also called the frictional term), and the anisotropic source term \citep{Lombriser:2015sxa, Lombriser:2016yzn, Sakstein:2017xjx, PhysRevLett.119.251301, Creminelli:2017sry, Ezquiaga:2017ekz,Nishizawa:2017nef, Belgacem:2017ihm, Belgacem:2018lbp, Belgacem:2019pkk,2020PhRvD.102d4009M}. Measurement of the GW signal from  astrophysical sources such as binary neutron stars (BNS), neutron star black holes (NS-BH), binary black holes (BBH) detectable from ground-based detectors (such as LIGO \citep{2015CQGra..32g4001L, PhysRevLett.123.231107, PhysRevLett.123.231108}, Virgo \citep{Acernese_2014}, and KAGRA \citep{Akutsu:2018axf}, and in the future from LIGO-India \citep{Unnikrishnan:2013qwa}, Cosmic Explorer\citep{Reitze:2019iox}, and Einstein Telescope \citep{Punturo:2010zz}, and the supermassive BBHs detectable from space-based GW detectors such as LISA \citep{PhysRevD.93.024003}, brings a unique way to test alternative theories of gravity via the propagation of GW through space-time over cosmological scales. Key aspects of alternative theories of gravity can be tested via GW propagation, and  cannot be tested via only EM observations. Our proposal provides a new window to test fundamental physics. 
 
Measurement of the electromagnetic counterpart within about $1.7$ seconds \citep{TheLIGOScientific:2017qsa, Abbott:2017xzu, Monitor:2017mdv} from the BNS event GW170817  {has enabled}  stringent constraints to be imposed on the speed of gravitational wave propagation and the mass of the graviton to high precision, and also constraints on the frictional term \citep{Abbott:2018lct, 2020PhRvD.102d4009M}. Recently, measurements of the plausible EM counterpart to the GW event GW190521 \citep{PhysRevLett.125.101102} by the  Zwicky Transient Facility (ZTF) collaboration at redshift $z=0.438$ \citep{PhysRevLett.124.251102} has enabled  constraints on the frictional term, though significantly weaker due to the large error ($\sim 50\%$) on the luminosity distance for GW190521 \citep{Mastrogiovanni:2020mvm}. Several forecast proposals have proposed  studying the frictional term via  GW detectors such as LIGO-Virgo \citep{2020PhRvD.102d4009M}, LISA \citep{Belgacem:2019pkk, Baker:2020apq}, and Einstein Telescope \citep{Belgacem:2018lbp, DAgostino:2019hvh, Hogg:2020ktc}  using the GW  sources that have EM counterparts. As a result, all of these studies are limited to only sources from which EM counterparts are expected, such as BNS, NS-BH, and  {supermassive black holes} (SMBHs) if there is a dedicated EM follow-up available. However, for most of the  BNS, NS-BH, and SMBHs, EM counterparts will not be detectable, and also for sources such as stellar mass BBHs, EM counterparts are unlikely if  baryonic matter is not present in their environment. For all such GW sources without EM counterparts, testing alternative theories of gravity is not possible in the current framework. Along with testing the propagation of GW signals, it is also possible to test other aspects of alternative theories of gravity \citep{PhysRevD.100.104036, Abbott:2020jks}   from the GW sources detectable from the network of LIGO-Virgo detectors \citep{LIGOScientific:2018mvr} \footnote{The catalog of the GW sources is available in this link \href{https://dcc.ligo.org/P2000061/public}{https://dcc.ligo.org/P2000061/public}}.  

Here we propose a new method that makes it possible to explore alternative theories of gravity from redshift-unknown GW sources (called dark sirens). Our method relies on exploiting the three-dimensional spatial clustering of the GW sources with  galaxy redshift surveys, as we  previously proposed for measuring the expansion history ($H_0$, $w_0$, $w_a$) and GW bias parameters ($b_{GW}(z)= b_{GW}(1+z)^\alpha$) \citep{Mukherjee:2018ebj, Mukherjee:2019wcg, Mukherjee:2020hyn}. Angular clustering to measure expansion history was proposed by \citep{PhysRevD.93.083511, Bera:2020jhx}. In this method, we propose to perform a cross-correlation of galaxy surveys with dark sirens to find the host redshifts of the dark sirens. Along with using the cross-correlation signal to find the host redshift, in our method, we propose to use the baryon acoustic oscillation (BAO) scale measured from the galaxy power spectrum to infer the angular diameter distance for the sources at that redshift. For metric theories of gravity, the angular diameter distance is uniquely related to the EM luminosity distance ($d_A(z)= d^{EM}_l (z)/(1+z)^2$). As a result, by combining the geometric distance measurements from BAO, and inferring the redshifts of the dark sirens using cross-correlation with galaxy surveys, we can measure the redshift dependence of the frictional term from the GW luminosity distance. 

The paper is organized as follows. In Sec. \ref{formalism}, we discuss the basic formalism of the method. In Sec. \ref{framework}, we discuss the statistical framework for implementation of the cross-correlation method. In Sec. \ref{mock} and Sec. \ref{results}, we discuss the set-up of the mock samples and forecasts for  measurement of the frictional term from the LIGO-Virgo-KAGRA detectors. In Sec. \ref{conc}, we conclude with the main findings of this method and discuss future prospects.

\section{Formalism: Relation between the geometric distance and the GW luminosity distance }\label{formalism}
GW propagation  in  space-time can be written according to the general theory of relativity as 
\begin{equation}\label{gw1}
    h_I'' + 2\mathcal{H}h_I' + c^2k^2h_I=0,
\end{equation}
where $h_I$ denotes the GW strain with  polarization states $I \in \{+, \times\}$ where the prime indicates the derivative with respect to the conformal time $\eta$, and $\mathcal{H}$ is the Hubble parameter in  comoving coordinates. However, in  alternative theories of gravity, GW propagation can be written as
\begin{equation}\label{modf1}
    h_I'' + 2(1-\gamma(z))\mathcal{H}h_I' + (c^2_{GW}k^2 +m^2_{GW}a^2)h_I=a^2\Pi_I,
\end{equation}
where $\gamma(z)$ is the frictional term, $c_{GW}$ is the speed of GW propagation, $m_{GW}$ is the graviton mass, and $\Pi_I$ is the anisotropic stress term. Comparing Eq. \ref{gw1} and Eq. \ref{modf1} indicates that for the general theory of relativity,  $\gamma(z)=0$, $c_{GW}=c$, $m_{GW}=0$, and $\Pi_I=0$.  Recent measurements from GW170817 have obtained strong constraints on $c_{GW}= c$, and $m_{GW}=0$ \citep{Monitor:2017mdv}. In the absence of the anisotropic source term $\Pi_I=0$, the effect of the frictional term $\gamma(z)$ leads to a modified luminosity distance to the GW source situated at redshift $z$ by the relation
\begin{equation}\label{modf2}
  d^{GW}_l(z)= \exp{\bigg(-\int dz' \frac{\gamma(z')}{1+z'}\bigg)}d^{EM}_l(z),
\end{equation}
where, $d_l^{EM} (z)$ is the luminosity distance according to the propagation of electromagnetic waves and which is related to the expansion history $H(z)$ by the relation $d_l^{EM}(z)= c(1+z)\int_0^{z} \frac{dz'}{H(z')}$, where $H(z)=H_0\sqrt{\Omega_m(1+z)^3 + (1-\Omega_m)}$ is related to  cosmological parameters such as the Hubble constant $H_0$, and the matter density $\Omega_m$ for the flat Lambda Cold Dark Matter (LCDM) cosmological model.
The above equation (Eq. \ref{modf2}) shows that the modification in the luminosity distance for the  GW can be larger or smaller than the EM luminosity distance for different theories of gravity. The parameter $\gamma(z)$ is degenerate with the electromagnetic (EM) luminosity distance $d_l^{EM}(z)$, and hence with other cosmological parameters. An independent probe of the EM luminosity distance from  electromagnetic observables can be useful for breaking the degeneracy between the cosmological parameters and the parameters related to  alternative theories of gravity $\gamma(z)$. 

One of the independent measures to the EM luminosity distance $d_l^{EM} (z)$ for a metric theory is through its relation with the geometric distance (the angular diameter distance) $d_A(z)= d_l^{EM}(z)/(1+z)^2$,  {according to  Etherington's reciprocity theorem \citep{2007GReGr..39.1055E} or the distance duality relation. This relation is valid for  EM probes if photon number is conserved, and photons propagate along null geodesics. The distance duality relation has been tested from several observations \citep{Holanda:2010vb, Holanda:2012at,Liao:2015uzb} and will also be tested more stringently in the future \citep{Liao:2015uzb, Renzi:2020bvl, Martinelli:2020hud, Arjona:2020axn}}.  The angular diameter distance to any redshift $z$ is related to the Baryon Acoustic Oscillation (BAO) scale $\theta_{BAO}$ in the matter correlation function by \citep{1970ApJ...162..815P,1984ApJ...285L..45B, Hu:1995en, Eisenstein:1997ik,Eisenstein:1997jh}
 \begin{equation}\label{bao}
\theta_{BAO} (z)= \frac{r_s}{(1+z)d_A(z)},
\end{equation}
where, $r_s= \int_{z_d}^\infty dz c_s(z)/ H(z)$ is the sound horizon where $z_d$ denotes the drag redshift. As a result, we can relate the BAO scale with the EM  luminosity distance $d_l^{EM}$ by the relation 
 \begin{equation}\label{bao2}
d_l^{EM} (z)= \frac{(1+z) r_s}{\theta_{BAO}(z)}.
\end{equation}
Using Eq. \ref{bao2} in Eq. \ref{modf2}, we can write 
\begin{equation}\label{modfandbao}
  d^{GW}_l(z)= \exp{\bigg(-\int dz' \frac{\gamma(z')}{1+z'}\bigg)}\frac{(1+z) r_s}{\theta_{BAO} (z)}.
\end{equation}
This is the key equation of this paper. In this expression, the measurement of the term $\frac{r_s}{\theta_{BAO}(z)}$ comes from  EM probes such as  large-scale structure galaxy redshift surveys \citep{Eisenstein:2005su, 2013AJ....145...10D, Alam:2016hwk} and CMB \citep{Spergel:2003cb, Spergel:2006hy,2011ApJS..192...18K, 2013ApJS..208...19H, Ade:2015xua, Aghanim:2018eyx}, and the measurement of $d_l^{GW}$ arises from the GW strain. We can write the above equation as 
\begin{equation}\label{modfandbao2}
  d^{GW}_l(z)\theta_{BAO}(z)= \exp{\bigg(-\int dz' \frac{\gamma(z')}{1+z'}\bigg)}(1+z) r_s.
\end{equation}
This relation shows that the product of the BAO angular scale $\theta_{BAO}(z)$ and the luminosity distance $d^{GW}_l(z)$ can measure the frictional term $\gamma(z)$ as a function of redshift. So, the concordance between the EM geometric probes and the GW luminosity probes allows a  way to test the theory of gravity. If the general theory of relativity is the correct theory of gravity, then the product between  $\theta_{BAO}(z)$ and $d^{GW}_l(z)$ should vary with redshift as $(1+z)$. Any deviation from this scaling can be a signature of alternative theories of gravity. The quantities $\theta_{BAO}$ and $d_l^{GW}$ are measured from large-scale structure and GW data, and the value of $r_s$ depends on  recombination physics and the sound speed in the baryon-photon fluid at the time of decoupling at redshift $z\approx 1100$ \citep{1968ApJ...151..459S, 1970ApJ...162..815P, 1970Ap&SS...7....3S, Hu:1995kot}. As a  result, this relation is nearly model-independent, and can be written directly in terms of  observables such as $d_l^{GW}(z)$, and $\theta_{BAO}(z)$. 

The BAO scale $\theta_{BAO} (z)$ can be inferred from the correlation function of the large-scale structure density field $\xi(r)$, and $r_s$ can be constrained from the cosmic microwave background (CMB) observations. As a result, the right-hand side of Eq. \ref{modfandbao} can be measured independently from large-scale structure observations and from CMB data. The inference of  $\theta_{BAO}(z)$ from  large-scale structure observations at redshift $z$, and the measurement of the GW luminosity distance from  GW sources situated at the same redshift $z$, will make it possible to reconstruct the frictional term $\gamma(z)$ as a function of redshift. The value of the sound horizon $r_s$ is obtained from  CMB measurements. Currently, the measurement of the CMB temperature and the polarization field from CMB experiments provides a measurement of the value of $r_s \approx 147$ Mpc \citep{Alam:2016hwk}.  

However, for this method to work, we need to infer the redshifts of the galaxies and also the redshifts of GW sources. The redshifts of the galaxies can be identified from  photometric or spectroscopic surveys. Several ongoing/upcoming missions such as eBOSS \citep{Alam:2020sor}, Dark Energy Survey (DES) \citep{10.1093/mnras/stw641}, Dark Energy Spectroscopic Instrument (DESI) \citep{Aghamousa:2016zmz}, Euclid \citep{2010arXiv1001.0061R}, Nancy Grace Roman Telescope \citep{2012arXiv1208.4012G, 2013arXiv1305.5425S, Dore:2018smn}, Vera Rubin Observatory \citep{2009arXiv0912.0201L}, Spectro-Photometer for the History of the Universe, Epoch of Reionization, and Ices Explorer (SPHEREx) \citep{Dore:2018kgp}, Subaru Prime Focus Spectrograph \citep{2014PASJ...66R...1T} will be covering nearly the full-sky up to  {redshift $z=3.0$}. The redshifts to the GW sources having EM counterparts (bright standard sirens) can be measured by identifying the host galaxy and the corresponding redshift from the catalog. For GW sources without EM counterparts (dark standard sirens), host galaxy identification is not possible. Dark standard sirens are expected to be detectable from larger luminosity distances (so up to high redshift) and hence should be more numerous due to the volume factor. Also, the redshift dependence of the frictional term can be measured from the sources that are distributed up to high redshift. So, it is important to be able to use dark standard sirens to reconstruct the frictional term up to high redshift. We discuss below a framework that can be used to infer the redshift using  cross-correlation with redshift-known galaxies.

 In this paper, we will consider a parametric form of the modification of the GW luminosity distance in terms of $\Xi_0$ and $n$ \citep{Belgacem:2017ihm, Belgacem:2018lbp, Belgacem:2019pkk}
\begin{equation}\label{modf3}
 \frac{d^{GW}_l(z)}{d^{EM}_l(z)}= \Xi_0 + \frac{1-\Xi_0}{(1+z)^n}.
 \end{equation}
 In terms of these parameters $\Xi_0$ and $n$, $\gamma(z)$ can be expressed by the relation 
 \begin{equation}\label{modf4}
 \gamma(z)=  \frac{n(1-\Xi_0)}{1-\Xi_0 + \Xi_0(1+z)^n}.
\end{equation}
 $\Xi_0=1$ and $n=0$ represents the fiducial value of these two parameters for the general theory of relativity. Our method can capture any functional form of $\gamma(z)$, and is not only restricted to this particular form. 

\section{Framework for testing GR from dark standard sirens}\label{framework}
To measure the frictional term using GW sources, it is important to infer the redshifts to the GW sources. For GW sources without EM counterparts, one cannot identify the host galaxy and its redshift. So we propose to use the cross-correlation with galaxy surveys to find the host redshift of the GW sources, as proposed by \citet{Mukherjee:2018ebj, Mukherjee:2019wcg, Mukherjee:2020hyn}. The auto-power spectrum $P^{ij}_{XX} (k,z)$ and cross power spectrum $P^{ij}_{XY} (k,z)$ between GW sources and galaxy samples can be written in terms of the matter power spectrum $P_m(k, z)$ by the relation \citep{Mukherjee:2019wcg, Mukherjee:2020hyn}
\begin{eqnarray}
     P^{ss}_{gg} (\vec k,z)&= b^2_g(k,z)(1 + \beta_g \mu_{\hat k}^2)^2P_{m}(k,z),\\\nonumber
     P^{sr}_{g\,GW} (\vec k,z)&= b_g(k,z)b_{GW}(k,z)(1 + \beta_g \mu_{\hat k}^2)P_{m}(k,z),\\\nonumber
     P^{rr}_{GW\,GW} (\vec k,z)&= b^2_{GW}(k,z)P_{m}(k,z), 
\end{eqnarray}
where $b_g (k,z)$ is the galaxy bias, $b_{GW} (k,z)$ is the GW bias parameter, $\beta_g=f/b_g$ which is related to the growth function $D$ by the relation $f\equiv \frac{d\ln D}{d \ln a}$ \citep{1980lssu.book.....P}. The term $\mu_k \equiv \cos(\hat n. \hat k)$ denotes the angle between the line of sight $\hat n$ and the Fourier modes $\hat k$. 

The auto-correlation between the galaxy samples at each tomographic redshift bins $P^{ss}_{gg} (\vec k,z)$ is a measure of the BAO scale $\hat \theta_{BAO}$. The convenient approach for this is to use the angular correlation function   
\begin{equation}
    \xi(\theta_{12}, \bar z)= \int dz_1 \Phi(z_1) \int dz_2 \Phi(z_2) \int \frac{dk k^2}{2\pi^2} b_g^2P_m(k, \bar z) j_0(kx)
\end{equation}
where $\Phi (z)$ is the selection function for the galaxies, $j_0(kr)$ is the zeroth order Bessel function, $\bar z = (z_1 +z_2)/2$ is the mean redshift, and $x= \sqrt{r(z_1)^2 +r(z_2)^2 -2 r(z_1)r(z_2)\cos(\theta_{12})}$ is the comoving distance between a galaxy pair denoted in terms of the angular separation between the galaxies $\theta_{12}$. The BAO scale can be obtained by fitting the angular correlation function with a power-law and a Gaussian model given by \citep{2012MNRAS.427.2146X, Carvalho:2015ica, Alam:2016hwk}
\begin{equation}
    \xi(\theta_{12}, \bar z)= A_1+ A_2\theta_{12}^k + A_3\exp\bigg(-\frac{(\theta_{12}-\theta_{FIT} (\bar z))^2}{\sigma_\theta^2}\bigg),
\end{equation}
where $A_i$ are the coefficients, $\sigma_\theta$ is the width of the BAO feature, and $\theta_{FIT}$ is related to the $\theta_{BAO}$ by the relation
\begin{equation}
   \theta_{BAO}(\bar z)= \theta_{FIT}(\bar z)+ \epsilon(\bar z, \Delta z)\theta_E(\bar z, \Delta z=0),
\end{equation}
where the second term arises due to the finite bin width $\Delta z= z_2- z_1$ and the correction term  $\epsilon \equiv 1- \theta_E(\bar z, \Delta z)/\theta_E(\bar z, \Delta z=0)$ is related to the change in the peak position $\theta_E(\bar z, \Delta z=0)$ in the limit of zero bin-width $\Delta z=0$. One can also use the linear point measurement to measure the BAO scale \citep{Anselmi:2017cuq}. 

 As proposed in previous work \citep{Mukherjee:2018ebj, Mukherjee:2020hyn}, the cross-correlation power spectrum between  GW sources and galaxy surveys $P^{sr}_{g\, GW} (\vec k,z)$  can be used to infer the host redshift shell of the GW sources. A similar approach to use the clustering signature to measure the Hubble constant has been  performed \citep{Bera:2020jhx} which depends on the method used \citep{PhysRevD.93.083511}. 
Using Bayes' theorem, we can write the joint estimation of the cosmological parameters $\Theta_c \in \{H_0, \Omega_m, w_0, w_a\}$, non-GR parameters $\Theta_{GR} \in \{\Xi_0, n\}$, and  bias parameters $\Theta_n \in \{b_{GW}, \alpha\}$ as 

\begin{eqnarray*}\label{posterior-1}
     \mathcal{P}(\Theta_{GR},\Theta_c, \Theta_n|\vec{\vartheta}_{GW}, \vec d_{g})\propto  \Pi(\Theta_c)\Pi(\Theta_{GR})  \, \\ \nonumber \times\, \Pi(\Theta_n)  \int \int dr_s \, dz    \bigg[\prod_{i=1}^{N_{GW}}\, \, \mathcal{L}(\vec{\vartheta}_{GW}| P^{ss}_{gg}(\vec k,z), \Theta_n, \vec d_g(z)) \\ \nonumber  \times \mathcal{P}(\vec d_g| P^{ss}_{gg}(\vec k,z)) 
     \mathcal{P}({\{d^i_l\}}_{GW}|z, \Theta_{GR}, \Theta_c, r_s, \{\theta^i,\, \phi^i\}_{GW})\\ \nonumber \times \mathcal{P}({\{\hat \theta_{BAO}\}}|\Theta_c,z) 
     \Pi(z) \Pi(r_s)\bigg] ,
\end{eqnarray*}
where each term can be written as follows:  $\Pi(X)$ denotes the prior on a quantity $X$,  $\mathcal{P}({\{\hat \theta_{BAO}\}}|\Theta_c,z)$ is the posterior on the BAO peak position given the cosmological parameters and the redshift,  $\mathcal{P}({\{d^i_l\}}_{GW}|z, \Theta_{GR}, \Theta_c, r_s, \{\theta^i,\, \phi^i\}_{GW})$ is the posterior on the luminosity distance given the cosmological parameters, GR parameters, redshift, sound horizon, and sky localisation of the GW sources, $\mathcal{P}(\vec d_g| P^{ss}_{gg}(\vec k,z))$ denotes the posterior on the galaxy power spectrum, and $ \mathcal{L}(\vec{\vartheta}_{GW}| P^{ss}_{gg}(\vec k,z), \Theta_n, \vec d_g(z))$ denotes the likelihood to estimate the source redshift of the GW sources, and can be written as 
\begin{eqnarray}\label{like-1}
&\mathcal{L}(\vec{\vartheta}_{GW}| P^{ss}_{gg}(\vec k,z), \Theta_n, \vec d_g(z)) \propto \exp\bigg(-\frac{V_s}{4\pi^2}\int k^2 dk \\ \nonumber  &\times \int d\mu_k  {\frac{ (\hat P (\vec k, \Delta \Omega_{GW}) - \tilde P(k,z))^2}{2(P^{ss}_{gg}(\vec k,z) + n_g(z)^{-1})(P^{rr}_{GW\,GW}(\vec k,z) + n_{GW}(z)^{-1})}}\bigg),
\end{eqnarray}
where $\hat P(\vec k, z)= \delta_{g}(\vec k, z)\delta_{GW}^*(\vec k,\Delta \Omega_{GW})$, $\tilde P(k,z)= b_g(k,z)b_{GW}(k, z)(1 + \beta_g \mu_{\hat k}^2)P_{m}(k,z)e^{-\frac{k^2}{k^2_{\rm eff}}}$, $n_{GW}(z)= N_{GW}(d^i_l(z))/V_s$ denotes the number density of gravitational wave sources in terms of  GW sources in the luminosity distance bin $N_{GW}(d^i_l(z))$, $\Delta \Omega_{GW}$ is the sky localisation error, and $V_s$ is the overlapping sky volume accessible to both GW and galaxy surveys. 

\section{Applying the method to GW and galaxy mock samples}\label{mock}
Galaxy samples: In this analysis, we generate the mock sample galaxies using the \texttt{nbodykit} \citep{Hand:2017pqn} with  box size with $1350\, \rm{Mpc/h}$ in the direction perpendicular to the line of sight, and the line-of-sight direction is considered to be  redshift $z$. The galaxy samples considered in this analysis are distributed up to redshift $z=1$. The mock sample is generated with a matter power spectrum $P_m(k)$ and using the cosmological parameters according to Planck-2015 \citep{Ade:2015xua}. We include redshift space-distortions in the mock sample according to the method prescribed in \citep{Hand:2017pqn}. For the mock samples, we consider the galaxy bias parameter $b_g= 1.6$ \citep{2012MNRAS.427.3435A, Desjacques:2016bnm,Alam:2016hwk}. The contribution from weak lensing is negligible for  sources below redshift $z=1$, and we have not included the contribution from weak lensing in this analysis. 

GW samples: GW mock samples are produced which follow the same underlying galaxy distribution over the redshift range $z=0.1$ to $z=1.0$. For the mock samples, we consider equal mass GW sources with each mass $30 M_\odot$ with two different sky localization error $10$ sq. deg and $100$ sq. deg, as expected from the network of the LIGO-Virgo-KAGRA detectors \citep{Chan:2018fpv}. The uncertainty in the luminosity distance error is calculated using the matched filtering technique  \citep{Sathyaprakash:1991mt,Cutler:1994ys,Balasubramanian:1995bm, 2010ApJ...725..496N} up to $f_{max}= f_{merg}$ which can be written in terms of the symmetric mass ratio $\eta= m_1m_2/ (m_1+m_2)^2$ and the total mass $M= m_1+m_2$ as $f_{merg}= c^3(a_1\eta^2 + a_2\eta +a_3)/\pi G M$, where  $a_1= 0.29740$,  $a_2=0.044810$, $a_3=0.095560$ \citep{Ajith:2007kx}. 
\begin{equation}\label{snr}
    \rho^2\equiv 4\int_0^{f_{max}} df \frac{ |h(f)|^2}{S_n(f)},
\end{equation}
where $\rho$ denotes the matched-filtering 
signal-to-noise ratio, $S_n(f)$ is the noise power spectrum for the LIGO design sensitivity \citep{Martynov:2016fzi} \footnote{The noise power spectrum is available on the webpage  \href{https://dcc-lho.ligo.org/LIGO-T2000012/public}{https://dcc-lho.ligo.org/LIGO-T2000012/public}} and the strain of the GW signal $h(f)$ can be written in terms of the redshifted chirp mass $\mathcal{M}_z= (1+z)\mathcal{M}_c$, inclination angle with respect to the orbital angular momentum $\hat L.\hat n$ (which is denoted by the function $\mathcal{I}_{\pm} (\hat L.\hat n)$), and luminosity distance to the source $d_L$ by the relation \citep{1987thyg.book.....H, Cutler:1994ys,Poisson:1995ef,maggiore2008gravitational, Ajith:2007kx}
 \begin{equation}\label{strain}
     h_{\pm}(f)= \sqrt{\frac{5}{96}}\frac{G^{5/6}\mathcal{M}_z^2 (f_z\mathcal{M}_z)^{-7/6}}{c^{3/2}\pi^{2/3}d_L}\mathcal{I}_{\pm} (\hat L.\hat n).
 \end{equation}
 
 For these simulations, we have considered our fiducial model to be  the LCDM model of cosmology with  parameter values according to Planck-2018 \citep{Aghanim:2018eyx},  { consistent with the Planck-2015 results \citep{Ade:2015xua}, and} the fiducial frictional terms ($\Xi_0=1$, and $n=0$) according to the general theory of relativity.  
 
\begin{figure*}
    \centering
    \includegraphics[width=1\linewidth]{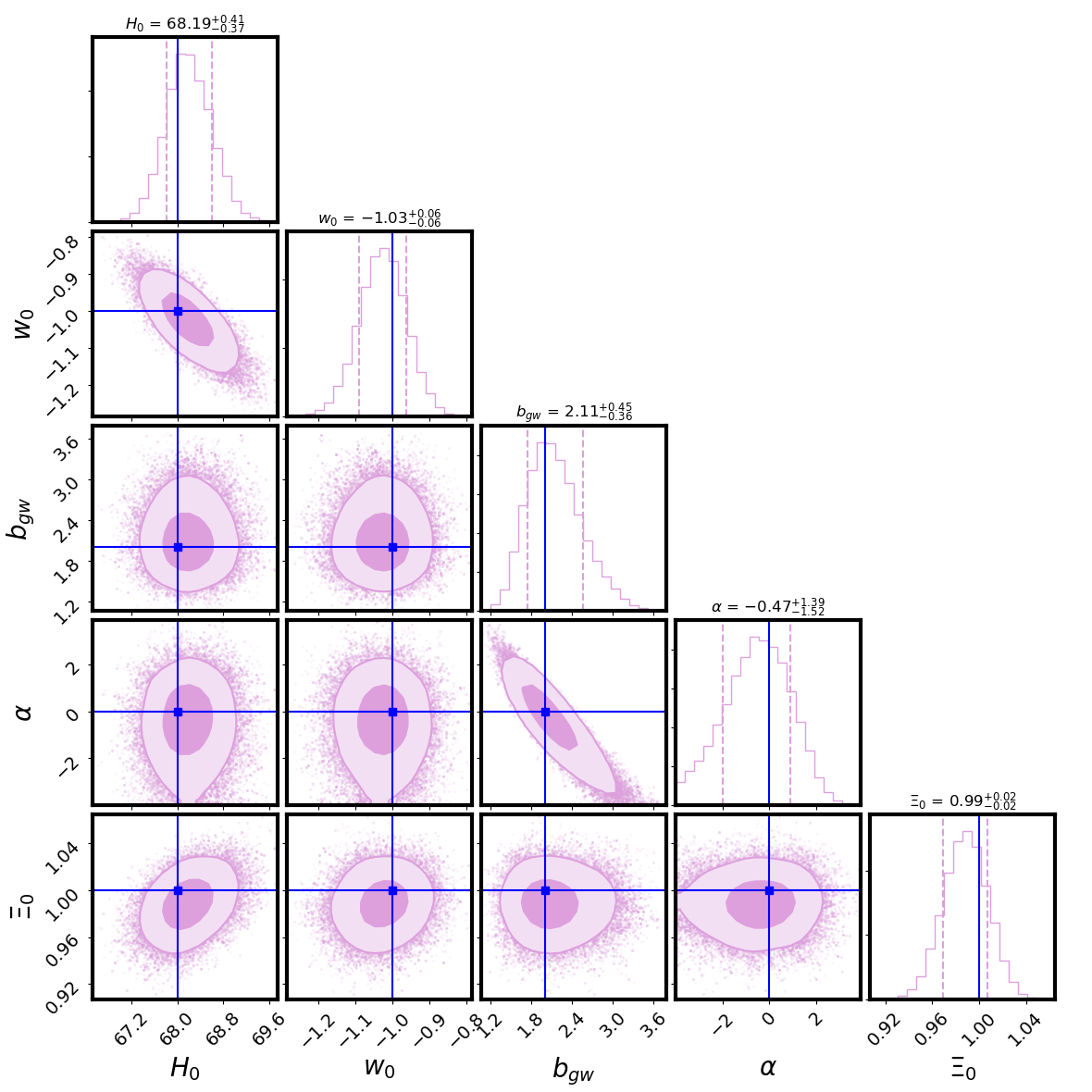}
    \caption{The joint estimation of the cosmological parameters $H_0, w_0$, GW bias parameters $b_{GW}, \alpha$, and non-GR parameter $\Xi_0$ with the fixed value of $n=1.5$. The number of GW sources is considered to be  $\sim 3500$ and detectable up to redshift $z=0.5$ with the aLIGO design sensitivity and with a sky localization error $\Delta \Omega_{GW}= 100$ sq. deg.  {The fiducial values used in the mock samples are shown by the blue solid line.}}
    \label{fig:withoutn0}
\end{figure*}

 \begin{figure*}
    \centering
    \includegraphics[width=1\linewidth]{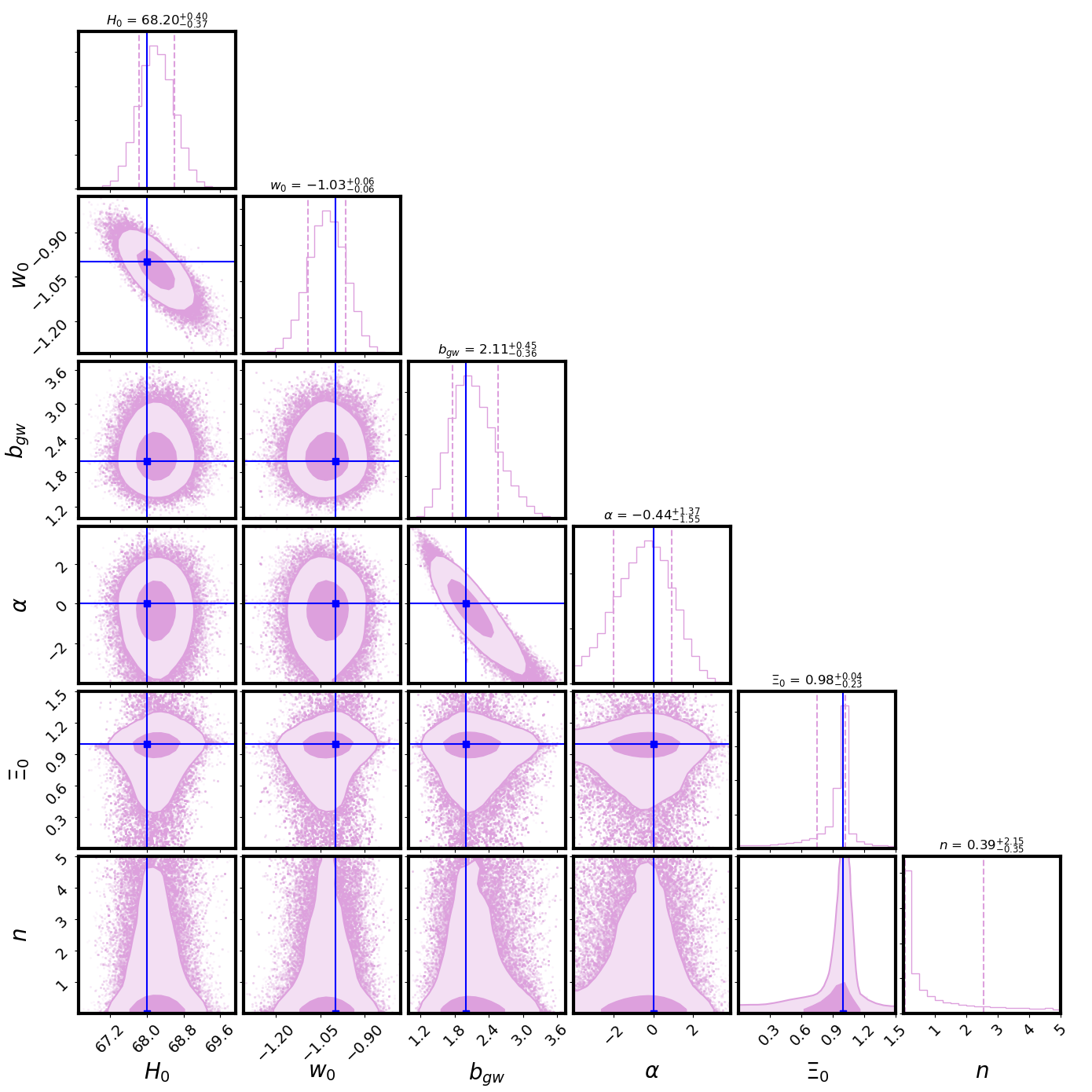}
    \caption{The joint estimation of the cosmological parameters $H_0, w_0$, GW bias parameters $b_{GW}, \alpha$, and non-GR parameters $\Xi_0,\, n$ with $\sim 3500$ GW sources detectable up to redshift $z=0.5$ at the aLIGO design sensitivity with a sky localization error $\Omega_{GW}= 100$ sq. deg.  {The fiducial values used in the mock samples are shown by the blue solid line.}}
    \label{fig:withn0}
\end{figure*}

\section{Forecast for the network of LIGO-Virgo-KAGRA detectors}\label{results}
In this analysis, we consider the combination of cosmological parameters ($H_0,\, w_0$), GW bias parameter, and its redshift dependence which is parameterized as a power-law $b_{GW}(z)= b_{GW}(1+z)^{\alpha}$ with two parameters $b_{GW}$ and $\alpha$, and the parameters related to the redshift dependence of the frictional term which can be parameterized in terms of $\Xi_0$, and $n$ (using Eq. \ref{modf3}).  The dependence of the parameters $\Xi_0$ and $n$ on redshift makes it essential for joint estimation of both the parameters in order to  not keep the redshift dependence at a fixed value, unlike several previous analyses \citep{Belgacem:2018lbp, Baker:2020apq}. The cross-correlation measurement of the GW sources with galaxies requires us to marginalize over the redshift-dependent GW bias parameter which depends on the GW source merger rate and population. We show the joint estimation for two cases, (i) five parameter model ($H_0, w_0, b_{GW}, \alpha, \Xi_0$) with the value of $n$ kept fixed at 1.5 from our setup, and (ii) six parameter model ($H_0, w_0, b_{GW}, \alpha, \Xi_0, n$).   For the forecast study, the error on the BAO scale $\theta_{BAO}$ is considered according to  current large-scale structure surveys \citep{Alam:2016hwk}. The priors on the parameters are considered flat as follows: $\Pi(H_0) \in [20, 150] \rm{km/s/Mpc}, \Pi(w_0)\in [-3, -0.1], \Pi(b_{GW}) \in [0,6], \Pi(\alpha) \in [-4,4], \Pi(\Xi_0) \in [0, 1.5] $, and $n \in [0, 5]$. The value of $r_s$ is  accordingly taken  to be fixed at the value considered by the eBOSS analysis \citep{Alam:2016hwk}. 

 We obtain results for the five parameters ($H_0, w_0, b_{GW}, \alpha, \Xi_0$) keeping the redshift dependence fixed at a value $n=1.5$.   {The results are not sensitive to the choice of the value of $n$. We find similar unbiased estimations also for other values such as  $n=0$, and $n=2.5$. Previous analyses were carried out with a fixed value of $n=2.5$ \citep{Belgacem:2018lbp}.} The measurability of such a scenario from the dark standard sirens detectable from the LVK network of detectors is shown in Fig. \ref{fig:withoutn0} for  {$N_{GW}= 3502$ ($\sim 3500$)} BBHs distributed up to redshift $z=0.5$ with sky localization errors $\Delta \Omega_{GW}= 100$ sq. deg. The use of the BAO scale for every redshift improves the constraining power of the cosmological parameters $H_0, w_0$. As a result, measurement of the non-GR parameter $\Xi_0$ is possible with about $2\%$ accuracy. A similar accuracy (about a factor of two better, $\Delta \Xi_0= 0.008$ \citep{Belgacem:2018lbp}) is only possible for 1000 binary neutron star sources with EM counterparts from the Einstein Telescope. Recent work has shown that LVK detectors can obtain only a $10\%$ measurement of the frictional term with BNS \citep{Baker:2020apq} for a fixed redshift dependence, which is a factor of five weaker than constraints possible with our method using $\sim 3500$ BBHs. 
 With the feasibility of using dark standard sirens proposed by our method, we can measure the modification in the GW luminosity distance up to high redshift ($z= 0.5$) with numerous sources detectable due to the larger accessible volume, and from  sources such as BBHs which are intrinsically louder than the BNS events due to the mass dependence.  {Assuming that the BBH merger rate will be governed by the Madau-Dickinson star formation rate \citep{Madau:2014bja}, we expect to be able to measure about $250-600$ GW sources per year from the advanced LIGO design sensitivity depending upon the mass distribution of the GW sources \citep{Fishbach:2018edt}. So, within a time-scale of about six to fourteen years of observation time of LVK detectors, we will be able to measure the $\Xi_0$ parameter with  $2\%$ accuracy.}
 Hence on a time-scale shorter than previously expected, this method will enable us to test the frictional term from GW observations. These joint studies show that the three-dimensional cross-correlation technique makes it possible to reduce the degeneracy between the GW bias parameter (which is related to the GW merger rates, and its population) with the parameters related to cosmology and theories of gravity. This arises in particular from the three-dimensional correlation function that takes into account the shape of the correlation function \citep{Mukherjee:2020hyn}. The shape of the correlation function is not affected by the bias parameter but is affected by the cosmological parameters. As a result, the inference of the clustering redshift becomes more robust through the three-dimensional correlation function, as previously shown by \citep{Mukherjee:2020hyn}.   {For cases with sky localization error $\Delta \Omega_{GW}= 10$ sq. deg. the error-bar on the GW bias parameters improves by about a factor of two due to the better estimate of the three-dimensional clustering signal. However, the error bars on the cosmological parameters and the non-GR parameters do not change significantly. This is because the improvement in the error bars associated with the redshift estimation is marginal in comparison to the error bars associated with the GW luminosity distances.}  

 \begin{figure}
    \centering
    \includegraphics[width=1\linewidth]{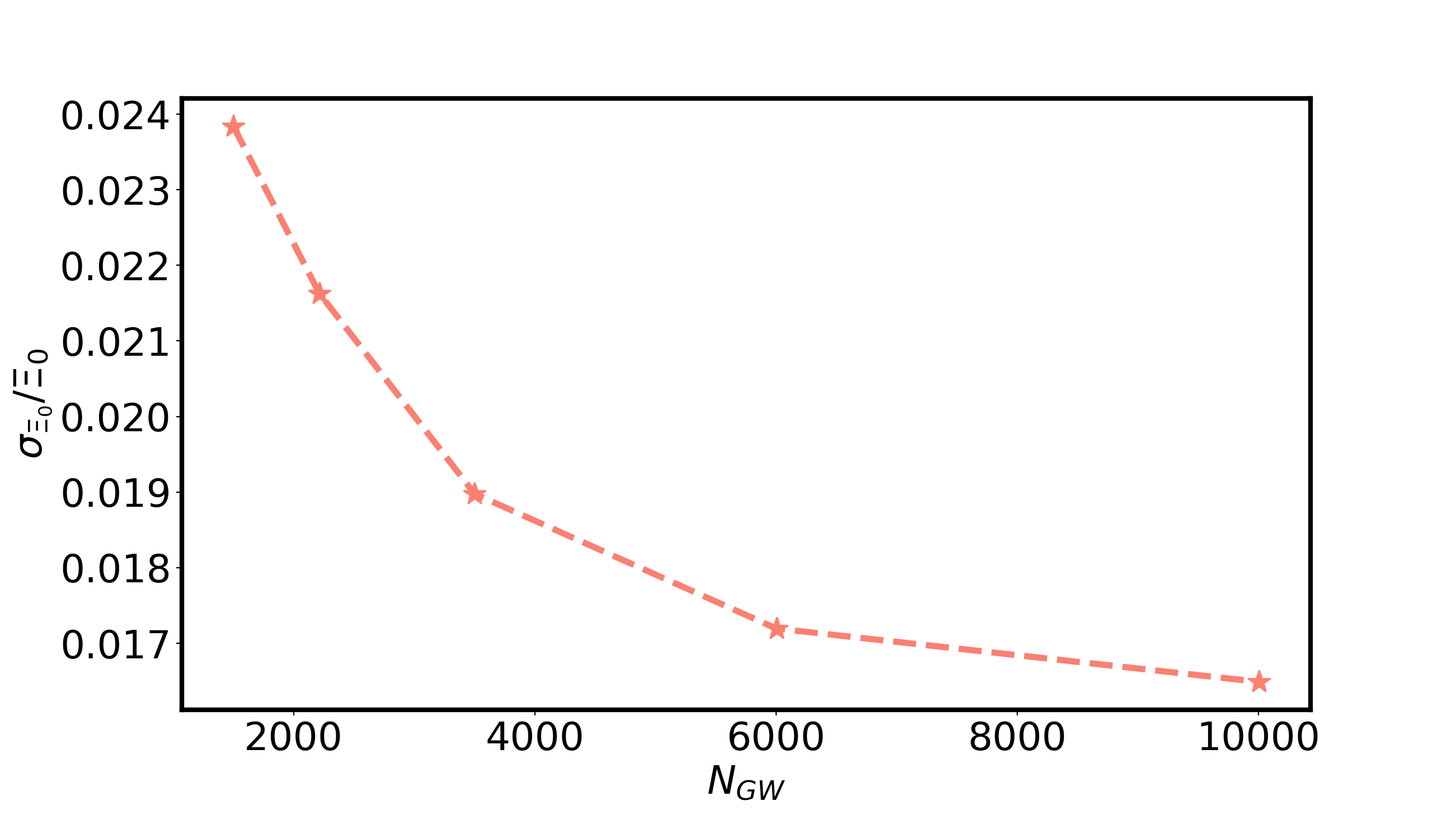}
    \caption{We show the change in the error-bar $\sigma_{\Xi_0}$ on the non-GR parameter $\Xi_0$ with change in the total number of GW sources $N_{GW}$ for sky localization error $100$ sq. deg after marginalizing over cosmological parameters $(H_0, w_0)$, GW bias parameters $(b_{GW}, \alpha)$, and fixed value of $n$.}
    \label{fig:comp}
\end{figure}

In Fig. \ref{fig:withn0} we show the forecast for the joint estimation of the six parameters ($H_0, w_0, b_{GW}, \alpha, \Xi_0, n$) for  sky localization error $100$ sq. deg for $\sim 3500$ GW sources distributed up to redshift $z=0.5$. The plot shows the existence of degeneracy between the non-GR parameters $\Xi_0$ and $n$. This indicates that keeping a fixed value of $n$ or assuming a particular form of the redshift dependence will significantly underestimate the error-bar. In comparison to the case with a fixed value of $n$, the posterior on $\Xi_0$ becomes non-Gaussian as can be seen from Fig. \ref{fig:withn0}. The values of $\Xi_0 \geq 1$ can be constrained with about $4\%$ accuracy (a factor of two degradation in comparison to the fixed $n$ case), and the values of $\Xi_0<1$ can be constrained with about $25\%$ accuracy, instead of $2\%$ possible from the case with fixed redshift dependence.  The large error in the GW luminosity distance at higher redshifts leads to poor constraints on the parameter $n$, and hence causes a broad tail in the posterior distribution of this parameter.  
Our results also show the existence of degeneracy between the non-GR parameters and the GW bias parameters and  other cosmological parameters.
The inclusion of BAO helps in breaking the degeneracy between the cosmological parameters ($H_0, w_0$) and the non-GR parameters ($\Xi_0, n$). 
Our 
method can be used to  jointly infer the cosmological parameters, bias parameters, and the non-GR parameters by
using the dark standard sirens. In Fig. \ref{fig:comp}, we show the uncertainty on the $\Xi_0$ parameter with the change in the number of objects from $1500$ to $10000$ for sources distributed up to redshift $z=0.5$ with sky localization error $\Delta \Omega_{GW}= 100$ sq. deg. The error on the frictional term scales roughly with the increase in the number of objects by $1/\sqrt{N_{GW}}$. The improvement in the measurement gets saturated (even with an increase in the number of GW sources), due to the presence of the error on the BAO scale. More precise measurement of the BAO scale will lead to  further improvement in the estimation of the non-GR parameters.

\section{Applicability to the dark standard sirens detectable from LISA and Cosmic Explorer/Einstein Telescope}\label{comparison}
The future space-based GW detector Laser Interferometer Space Antenna (LISA) \citep{2017arXiv170200786A} and ground-based GW detectors such as the Einstein Telescope \citep{Punturo:2010zz} and Cosmic Explorer \citep{Reitze:2019iox} are going to detect numerous GW sources beyond redshift $z=1$. However, several of these sources are not going to have EM counterparts, and hence the current techniques are limited to only those sources for which EM counterpart detection is possible. Though BNSs, NS-BHs, and SMBHs are expected to have electromagnetic counterparts, their detection from an EM follow-up telescope can be challenging due to fading EM counterparts, incomplete galaxy catalogs, large sky localization errors, and unavailability of EM telescopes with cadence. Along with the bright standard sirens, there are also going to be numerous dark standard sirens up to high redshift for which EM counterparts cannot be measured. 

The method proposed in this paper can be used for the dark standard sirens detectable from LISA and Einstein Telescope/Cosmic Explorer by exploring the cross-correlation of the spatial position of the GW sources with the multi-frequency EM data which can be detected from Euclid \citep{2010arXiv1001.0061R}, Vera Rubin Observatory \citep{2009arXiv0912.0201L}, SPHEREx \citep{Dore:2018kgp}, Nancy Grace Roman Telescope \citep{2012arXiv1208.4012G, 2013arXiv1305.5425S, Dore:2018smn} and SKA \citep{Maartens:2015mra}. The spatial cross-correlation of GW sources with  galaxies provides the clustering redshift to the GW sources, and by combining the measurement of the BAO scale from auto-correlation of the galaxies detectable from these surveys, we will able to measure the frictional term as a function of redshift. In future work \citep{Mukherjee:2020:new}, we will study the feasibility of this method for LISA, the Einstein Telescope, and Cosmic Explorer in synergy with the large-scale structure probes. Due to the availability of a large number of GW sources with better luminosity distance measurements from  future GW detectors, the error budget from the GW sector is going to be reduced. Also with the availability of greater numbers of galaxies from future 
large-scale structure surveys, it will be possible to measure the BAO scale to an accuracy $\leq 1\%$ in the future \citep{Zhan:2008jh, 2009arXiv0912.0201L}. The major source of error on the non-GR parameter $\Xi_0$ will be the error associated with the GW sources. With  spectroscopic measurements of  galaxy redshifts, the relative error $\Delta \Xi_0\equiv \sigma_{\Xi_0}/\Xi_0$ ($\sigma_{\Xi_0}$ denotes $1$-$\sigma$ error-bar on the non-GR parameter $\Xi_0$) will approximately scale with the relative error on the luminosity distance $ \Delta d_l \equiv \sigma_{d_l}/d_l$ and relative error $\Delta_{BAO} \equiv \sigma_{\theta_{BAO}}/\theta_{BAO}$ on the measurement of the BAO-scale by
\begin{eqnarray}
\Delta \Xi_0^2 &\sim  \bigg[\sum_{z_{bin}}\frac{1}{\frac{\Delta^2_{d_l}(z)}{N_{GW}(z)}+\Delta_{BAO}^2(z)}\bigg]^{-1},
\end{eqnarray}
where $z_{bin}$ denotes the number of redshift bins and $N_{GW}(z)$ denotes the number of GW sources at the redshift $z$. So, we expect to be able to provide more stringent constraints using this method for dark standard sirens than accessible from only those sources with EM counterparts in future, a large number of sources having better luminosity distance measurements than provided by current detectors. With the availability of a large number of GW sources, the contribution to the total error on $\Xi_0$ associated with the GW  measurement will become subdominant relative to  uncertainties arising from the measurement of the BAO scale. As a result, accurate and precise measurement of the $\Xi_0$ parameter will be limited by the error associated with the BAO scale, when the number of GW sources will be $N_{GW} (z) >> \Delta^2_{d_l} (z)/\Delta^2_{\theta_{BAO}}(z)$. 
With  future detectors such as LISA, Cosmic Explorer, and Einstein Telescope, testing of the theory of gravity will be possible not only from the frictional terms but also from the lensing of gravitational wave \citep{Mukherjee:2019wfw, Mukherjee:2019wcg}. In future work, we will show the constraints on different theories of gravity by combining both of these aspects. 

\section{Conclusion}\label{conc}
GW propagation  in space-time provides a unique way to test theories of gravity by measuring the frictional term. The non-zero value of the frictional term leads to modification in the luminosity distance to the GW source from the canonical expression which is probed by the EM observable. The success of this avenue to test the theory of gravity depends on two quantities (i) accurate redshift identification to the GW sources, (ii) identifying the EM distances to this redshift. In all the existing methods, EM counterparts are essential to measuring the frictional term. As a result, all these methods are limited to only  GW sources such as binary neutron stars, neutron star-black holes, and supermassive binary black holes, for which EM counterparts are likely to occur. Also, due to the limitations of observing the EM follow-ups, EM counterparts to all such GW sources are not going to be detectable, even though they exist. Also, most sources, such as the stellar mass BBHs which are currently detectable from the LVK detectors and are more numerous than the GW sources with EM counterparts, cannot be used to measure the frictional term in the currently existing method. So, the current methods only rely on sources at  low redshift $z\leq 0.1$ to measure the friction term from the LVK detectors. However, the variation of the luminosity distance for redshift $z>0.1$ is significant in several theories of gravity \citep{Belgacem:2017ihm, Belgacem:2019pkk}.
The current method fails to use the high redshift sources from the LVK detectors. 

In this paper, we propose a method that avoids the existing hurdle of using only bright GW sources to measure the frictional term. We show that by exploiting the three-dimensional clustering scale of the GW sources with galaxies, one can use  dark standard sirens up to high redshift to measure the redshift dependence of the frictional term. Along with the usage of the three-dimensional clustering between GW sources and galaxies, we propose to use the BAO scale determined from the galaxy correlation function which provides an independent measure of the geometric distance at any redshift as probed by the EM observations. By combining BAO measurements from the galaxies and luminosity distance measurements from  GW sources, and by using the cross-correlation between  galaxies and GW sources, we show that one can make a joint measurement of the cosmological parameters, non-GR parameters, and also the GW merger rates and population denoted by the GW bias parameters $b_{GW}(z)$.

We argue that with $\sim 3500$ dark standard sirens detectable from the LVK detectors up to redshift $z=0.5$, one will be able to measure the frictional term with an accuracy of about $4\%$ for $\Xi_0\geq 1$, and with about $25\%$ accuracy for $\Xi_0<1$, after marginalizing over the redshift dependence for binary black holes of masses $30\, \rm{M}_\odot$ each. However, when a fixed redshift dependence of the frictional term is assumed, then the error on the parameter $\Xi_0$ improves to $2\%$ for both ($\Xi_0 <1$, and $\Xi_0\geq 1$) for the same BBH masses and  redshift distribution. These results show that  measurement of the frictional term is possible from the dark sirens detectable with the LVK detectors, and goes beyond the sensitivity of  currently existing methods. Moreover, the accuracy possible by our method with $\sim 3500$ GW sources of mass $30\, \rm{M}_\odot$ is nearly comparable with $1000$ GW binaries with EM counterparts detectable from  future GW detectors such the Einstein Telescope \citep{Belgacem:2018lbp}, which may not be operational until 2040 or later. As a result, by using the cross-correlation technique proposed in this work, we can achieve a similar uncertainty on the frictional term from the currently ongoing GW detectors, and hence in a times-cale shorter than possible from ET sources with EM counterparts. This technique will also be useful for the dark standard sirens detectable from the  future space-based detector LISA and ground-based detectors Einstein Telescope and Cosmic Explorer.

\section*{Acknowledgement}
 Authors would like to thank the anonymous referee for providing very useful suggestions on the manuscript. Authors are also thankful to Simone Mastrogiovanni for reviewing the manuscript and providing useful comments. SM acknowledges useful discussion with Martin Hendry, Simone Mastrogiovanni, and Daniele Steer during the presentation in the cosmology working group of the LIGO Scientific Collaboration.  This work is part of the Delta ITP consortium, a program of the Netherlands Organisation for Scientific Research (NWO) that is funded by the Dutch Ministry of Education, Culture and Science (OCW). The work of BDW is supported by the Labex ILP (reference ANR-10-LABX-63) part of the Idex SUPER,  received financial state aid managed by the Agence Nationale de la Recherche, as part of the programme Investissements d'avenir under the reference ANR-11-IDEX-0004-02. The Center for Computational Astrophysics is supported by the Simons Foundation.  This analysis was carried out at the Horizon cluster hosted by Institut d'Astrophysique de Paris. We thank Stephane Rouberol for smoothly running the Horizon cluster. We acknowledge the use of following packages in this work: Corner \citep{corner}, emcee: The MCMC Hammer \citep{2013PASP..125..306F}, IPython \citep{PER-GRA:2007}, Matplotlib \citep{Hunter:2007},  nbodykit \citep{Hand:2017pqn}, NumPy \citep{2011CSE....13b..22V}, and SciPy \citep{scipy}. The authors would like to thank the  LIGO/Virgo scientific collaboration for providing the noise curves. LIGO is funded by the U.S. National Science Foundation. Virgo is funded by the French Centre National de Recherche Scientifique (CNRS), the Italian Istituto Nazionale della Fisica Nucleare (INFN), and the Dutch Nikhef, with contributions by Polish and Hungarian institutes.
 
 \section*{Data Availability}
The data underlying this article will be shared on a reasonable request to the corresponding author. 

\def\urlprefix{}
\def\url#1{}
\bibliography{main.bib}
\label{lastpage}

\label{lastpage}

\end{document}